\documentclass[12pt]{article}

\usepackage{amstext,amsmath,amssymb,amsfonts,bbm,slashed}
\usepackage[latin1]{inputenc}
\usepackage{epsfig}
\usepackage{hyperref}
\usepackage{amsthm}
\usepackage{subfigure}
\usepackage{color}
\usepackage{multirow}

\newcommand{\ext}{\text{ext\,}}

\newcommand{\cG}{{\mathcal G}}

\newcommand{\bG}{{\partial\mathcal G}}
\newcommand{\bGa}{{\partial{\mathcal G'}}}
\newcommand{\tJ}{{\widetilde{J}}}

\newcommand{\cexG}{\mathcal G_{\text{color}}}
\newcommand{\bJ}{ J_{\partial} }

\newtheorem{lemma}{Lemma}

\newtheorem{definition}{Definition}

\newtheorem{proposition}{Proposition}

\newcommand{\bea}{\begin{eqnarray}}
\newcommand{\eea}{\end{eqnarray}}
\newcommand{\beq}{\begin{equation}}
\newcommand{\eeq}{\end{equation}}

\makeatletter

\begin{document}

\begin{titlepage}
\begin{flushright}
pi-qg-296\\
Lpt-Orsay-12-95\\
ICMPA-MPA/2012/12
\end{flushright}

\vspace{20pt}

\begin{center}

{\Large\bf Addendum to ``A Renormalizable 4-Dimensional \\
\medskip
Tensor Field Theory''}
\vspace{15pt}

{\large Joseph Ben Geloun$^{a,c,\dag}$ and Vincent Rivasseau$^{b,a,\ddag} $}

\vspace{15pt}

$^{a}${\sl Perimeter Institute for Theoretical Physics}\\
{\sl 31 Caroline St. N., ON, N2L 2Y5, Waterloo, Canada}\\
\vspace{5pt}

$^{b}${\sl Laboratoire de Physique Th\'eorique, CNRS UMR 8627}\\
{\sl Universit\'e Paris-Sud, 91405 Orsay, France}
\vspace{5pt}

$^{c}${\sl International Chair in Mathematical Physics and Applications\\ (ICMPA-UNESCO Chair), University of Abomey-Calavi,\\
072B.P.50, Cotonou, Rep. of Benin}\\
\vspace{5pt}
E-mails:  {\sl $^{\dag}$jbengeloun@perimeterinstitute.ca, 
$^\ddag$rivass@th.u-psud.fr}

\vspace{10pt}

\begin{abstract}
This note fills a gap in the article with title above \cite{bgriv}. We provide
the proof of Equation (82) of Lemma 5 in \cite{bgriv} and thereby 
complete its power counting analysis with a more precise next-to-leading-order estimate.
\end{abstract}
\end{center}

\noindent  Pacs numbers:  11.10.Gh, 04.60.-m
\\

\setcounter{footnote}{0}

\today

\end{titlepage}

\section{Introduction}

Recently, a just renormalizable tensor quantum field  model in four dimensions was introduced and analyzed by the present authors \cite{bgriv}. 
This model has possible relevance for a quantum theory of gravity \cite{Riv}, since it effectuates in a new way a 
statistical sum over simplicial pseudo-manifolds in four dimensions. It has been subsequently proved asymptotically
free in the ultraviolet regime \cite{bg}. From the physical point of view, this hints at a likely phase
transition in the infrared regime.
 
The renormalization of the model followed from a multi-scale analysis and a generalized locality principle, leading to a power-counting theorem. 
The divergent graphs were identified, leading
to the list of all marginal and relevant interactions. But it escaped our attention that one inequality (Equation (82), see Lemma 5 in \cite{bgriv}) which had been used to establish this list (see Equation (85) Lemma 6 and the discussion after Equation (90)) had no proper
proof provided\footnote{We thank the anonymous referee of another related paper \cite{BenGeloun:2012pu} for remarks that 
lead us to detect this gap in \cite{bgriv}.}. 

In this addendum, we close this gap by 
providing a full-fledged proof of Equation (82) and, in fact, an improved
bound of the same form.  
Therefore \cite{bgriv} and all subsequent papers hold without change.

For this purpose, our new inequality, written in the notations of \cite{bgriv} Section 5, is given by:
\begin{proposition}\label{prop1}
The degree of a rank-4 uncolored tensor graph with $\sum_{\bJ} g_{\bJ} = q >0$ satisfies
\beq\label{ineq}
 \sum_{J} g_{\tJ} -4q \geq 6\,,
\eeq
which is the claim of claimed of Equation (82) of  \cite{bgriv}.
\end{proposition}

In the rest of this note, we establish some general lemmas valid for tensor graphs of any rank
and which hold jacket by jacket. Proposition \ref{prop1} is then deduced as a special case of 
a whole set of similar results that could be established with these lemmas.
In fact, this note is therefore also a first step in a possible future systematic study of  $1/N$-subdominant 
contributions in the wider context of general \emph{uncolored} \cite{BGRuncoloring} tensor models \cite{GurRyan} and tensor group field theories \cite{bgriv,Riv,bg,BenGeloun:2012pu,Geloun:2012bz,COR} 
(see also \cite{Gurau:2011sk} for related results in the colored context).

\section{Deletion and contraction moves on graphs}

In this section, we recall some facts about uncolored tensor graphs $\cG$
of rank $D$, which have $D$-stranded lines of color 0, and external half-lines (also of color 0). They
can be related by a one-to-one correspondence to $D+1$ colored tensor graphs $\cexG$ \cite{bgriv,BGRuncoloring,universal}  (see, in particular, Definition 1 in \cite{bgriv}), which 
have $D$-stranded lines of colors 0, 1, \dots $D$. The lines of color 1, \dots $D$ will be
called below the \emph{colored lines}, and the 0-lines will be also called \emph{white lines}.
We shall establish some properties of these graphs under contraction or deletion
of these white lines. 

\begin{figure}
 \centering
     \begin{minipage}[t]{.7\textwidth}
      \centering
\includegraphics[angle=0, width=3cm, height=2cm]{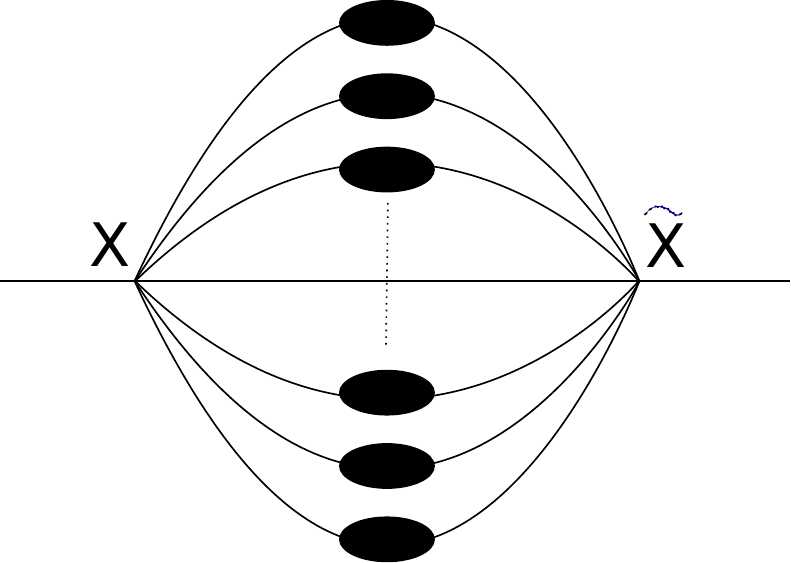}
\vspace{0.1cm}
\caption{ {\small A melon made with $D$ two-point functions with two mirror vertices 
${\rm x}$ and $\tilde {\rm x}$.}}
\label{fig:mirror}
\end{minipage}

\end{figure}

Let us recall that the dominant graphs of the tensor $1/N$ expansion are called melonic graphs or simply melons
\cite{GurRyan,melons}. In a melon, every vertex $x$ has a single mirror vertex $\tilde x$ such that these two vertices  are 
connected by $D$ two-point functions, each of a different color (see Fig. \ref{fig:mirror}).

The first lemma below is already known in the context of ribbon graphs and of colored tensor graphs \cite{GurRyan}
but not enunciated in the uncolored context 
(which is the relevant one here).   

We start by recalling the definition of tree contraction.
\begin{definition}

Given a connected (uncolored) tensor graph $\cG$ and a tree $T$ of white lines, 
contracting the tree leads to a reduced graph  $\cG_1 = \cG/T$ called a (tensor) ``rosette''\footnote{In \cite{Gurau:2011xq}, contracting
successively a tensor rosette, with respect to a tree of lines in each color,  yields the so-called  ``core graph''. In a sense, the contraction of a tree of white line can be called a ``pre'' or ``0-core'' graph. Inspired by the ribbon graph case, we use the name of rosette also in the tensor case.}. It has 
a single vertex which is a $D$-colored tensor graph plus white \emph{loop} lines and external lines
hooked to that vertex. 

\end{definition}

\begin{lemma}
\label{lem1}
We have, for any  connected $\cG$ and $T$, 
\beq
 g_{\tJ} = g_{\tJ_1} \,, \qquad  g_{\bJ} = g_{ {\bJ}_1} \,.
\eeq
\end{lemma}

\medskip
\noindent{\bf Remark 1.} 
We use the obvious notations where the subscript 1 means that the quantities are computed in the
tensor rosette graph  $\cG_1 = \cG/T$. The result therefore
holds for any pinched jacket $\tJ$ of $\cexG$ associated with $\cG$
yielding after contraction the jacket $\tJ_1$ of $\cG_{1} = \cG/T $, 
and any boundary jacket $\bJ$ of the boundary graph $\bG$ yielding 
after contraction the boundary jacket ${\bJ}_1$ of the boundary of $\cG_{1}$. 

\medskip
\noindent{\bf Proof of Lemma \ref{lem1}.} 
Let $V$ be the number of vertices,   
$L$ be the number of lines, $F$ the number of faces,
and $N_{\ext}$ the number of external legs of $\cexG$
extending $\cG$\footnote{ Note that in \cite{bgriv}, we use
slightly  different notations at this level:  $V$ is denoted by $V_{\cexG}$
and $L$ by $L_{\cexG}$. Nevertheless,  the present situation is totally 
unambiguous.}. 
We know that\footnote{ See comments in the proof of Theorem 2 in \cite{bgriv} leading to Equations (45), (48) and (49). We rewrite
these in the present context as Equation \eqref{f0kl}.} $F$ can be decomposed in 
\bea
F = \sum_{k} F_{0k} + \sum_{k} F^\ext_{0k} + \sum_{kl} F_{kl}\,,
\label{f0kl}
\eea
where $0,k,l$ refer to color indices such that 
$F_{0k}$ denotes the number internal faces of $\cG$
(of specific pair of colors $0k$), 
$F_{kl}$ the internal faces of $\cexG$ which does 
belong to $\cG$ (the colors $k,l \neq 0$), 
and $F^\ext_{0k}$ external  (open) faces of $\cG$. 

After a single tree  line  contraction (note that also this tree line contraction can be referred to as a 
dipole contraction with two different connected components 
at the end points of the dipole), we have
\beq
V \to V - 2 \,,  \qquad 
L \to L - (D+1) \,, \qquad 
N_{\ext} \to N_{\ext}\,,
\eeq
whereas for faces, one gets the modifications:
\bea
F_{0k} \to F_{0k} \, , \qquad 
F_{kl} \to F_{kl} - 1 \, , 
\qquad 
F^\ext_{0k} \to F^\ext_{0k}\,.
\eea
Let us now consider a jacket $\tJ=(n0m\dots)$ written as a cycle
of colors. 
Some alternating pairs of open faces $F^{\ext}_{0n}$
and $F^{\ext}_{0m}$ can merge into a single closed face; this happens when they meet white external legs through the ``pinching"
prescription. 
The total number of such merged faces is called $\widetilde F_{0nm}$
and so, since the jacket $\tJ$ is connected follows from the fact that
$\cG$ is connected and $V_{\tJ}=V$ and  $L_{\tJ}= L$, 
\bea
2 -2 g_{\tJ} = V -  L + F_{0n} + F_{0m}+ \widetilde  F_{0nm}+ \sum_{kl}F_{kl; \,\tJ}\,,
\eea
where the sum over $k,l$ is over $D-1$ (color) pairs in the jacket. 
Since every term in the sum in $k,l$ changes by $-1$ and
$V-L$ by $D-1$, and since $F_{0n}$, $F_{0n}$ and $\widetilde  F_{0nm}$
have not changed therefore, after contraction,
\bea
2 -2 g_{\tJ} \; \to \;2 -2 g_{\tJ}  \quad \Rightarrow\quad  g_{\tJ} = g_{\tJ_1}\,.
\eea
Consider, finally, a boundary jacket $\bJ$ in $\bG$,
$V_{\bG} = V_{\bJ} = N_{\ext}$, $L_{\bG} = L_{\bJ} = \sum_{k} F^\ext_{0k}  $ are constant. Then, in fact, $\bG$ is exactly the same. Hence
the genus of any boundary jacket $\bJ$ cannot change after contraction.
\qed

\medskip
\noindent
{\bf Remark 2.}  If every initial vertex of the graph $\cG$ is a $D$-melon (as is the case for the graphs considered
in \cite{bgriv}), then, for any tree $T$, the vertex of rosette $\cG_1$  is again a $D$-melon.

\begin{figure}
 \centering
     \begin{minipage}[t]{.7\textwidth}
      \centering
\includegraphics[angle=0, width=7cm, height=3cm]{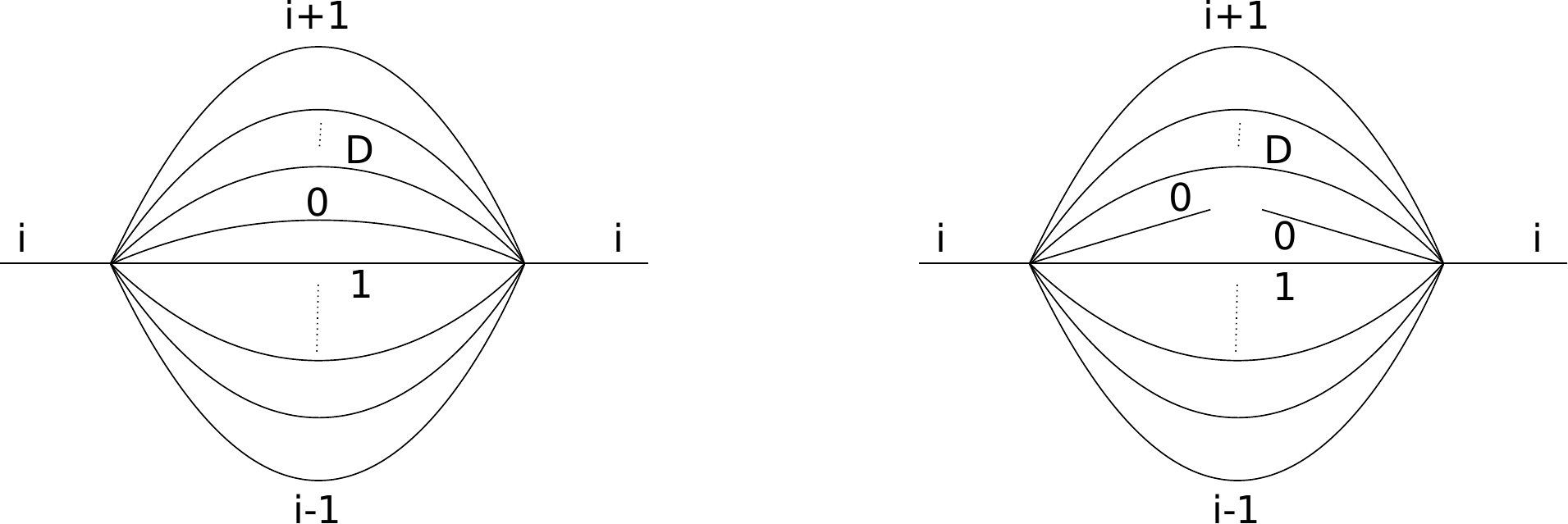}
\vspace{0.1cm}
\caption{ {\small Closed (left) and open (right) melopole  (elementary melon or $D$-dipole containing the color 0) of ``external'' color i.}}
\label{fig:melo}
\end{minipage}

\end{figure}

 \begin{definition}
Given a graph, we define a ``closed melopole'' as 
an elementary $D$-dipole made with a 0-line
and $D-1$ colored lines (see Fig. \ref{fig:melo}).

An  ``open melopole''
is defined in the same way but with the 0-line replaced
by two open external legs (of course of color 0) (see Fig. \ref{fig:melo}).

The color of a melopole is the missing color 
in the dipole defining it (it is also called its``external'' color,
see ``$i$'' in Fig. \ref{fig:melo}).

\end{definition}

\begin{definition}
If the single vertex of a rosette is a ($D$) melon graph, we say 
that the rosette is vertex-melonic. 

The melopole contraction of a graph is the recursive
contraction of all its closed and open melopoles until 
none are left. 

If the end result of the melopole contraction of a rosette
is such that no 0-lines are left, we say that the initial rosette is 0-melonic.

A rosette both vertex and 0-melonic is called fully-melonic. 

A melonic line of a vertex-melonic rosette is a line
joining two mirror vertices, otherwise it is called non-melonic. 

\end{definition}

Obviously, a vertex-melonic rosette which is not fully-melonic
must have at least one non-melonic line. 

\medskip
\noindent
{\bf Remark 3.} The result of the melopole contraction of a graph 
is independent of the order chosen to perform the recursive contraction of its open and closed melopoles. 
Given a graph $\cG$, we denote $\cG/M$ the graph reduced by the melopole contraction. 

In the same previous notations, the following statement holds:
\begin{lemma}\label{lem2}
The melopole contraction of any graph does not change the genera of its internal and boundary jackets:
\bea
&&
\forall \tJ\,, \qquad g_{\tJ} = g_{\tJ_1} \,,\crcr
&&
\forall \bJ\,, \qquad g_{\bJ} = g_{ {\bJ}_1} \,.
\eea
\end{lemma}
\noindent{\bf Proof.}
Let us consider first the contraction of a closed melopole of color
$i$. We have
\beq
V \to V - 2 \,,  \qquad 
L \to L - (D+1) \,, \qquad 
N_{\ext} \to N_{\ext}\,. 
\eeq
In the same notations of \eqref{f0kl}, it can be seen that
\bea
&&
F_{0k} \to F_{0k}-1 \, ,\; k\neq i,  \qquad
F_{0i} \to F_{0i} \, , \crcr
&&
F_{kl} \to F_{kl} - 1 \, , \; k,l\neq i,  \qquad
F_{ki} \to F_{ki} \, , \crcr
&&
F^\ext_{0k} \to F^\ext_{0k}\,, \quad \forall k\,.
\label{fkl1}
\eea
Choose a jacket $\tJ=(n0m\dots)$. 
There are two pairs in the jacket containing
$i$ whose face number does not change
and so, from \eqref{fkl1}, $D-1$ remaining change by -1. 
We infer that $g_{\tJ}$ does not change. 
Moreover the boundary graph has not changed since
$F^\ext_{0k} \to F^\ext_{0k}$. 

Next, let us consider the case of contraction of an open 
melopole of color $i$. We note the following:
\beq
V \to V - 2 \,,  \qquad 
L \to L - D \,, \qquad 
N_{\ext} \to N_{\ext}-2\,,
\eeq
as well as 
\bea
&&
F_{0k} \to F_{0k}\, ,\quad k \neq i\,, \qquad
F_{0i} \to F_{0i} \;\; \text{or}\;\; F_{0i} \to F_{0i} +1\, ,
\qquad 
\crcr
&&
F_{kl} \to F_{kl} - 1 \, , \; k,l\neq i,  \qquad
F_{ki} \to F_{ki} \, , \crcr
&&
F^\ext_{0k} \to F^\ext_{0k}-1\,.
\eea
Let $\tJ=(n0m\dots)$ be a jacket. Two cases might occur:

\begin{enumerate}
\item In a first subcase $i\neq n,m$: $D-3$ faces of the type $kl\neq 0i$, 
jump by -1; $\widetilde F_{0nm}$ also jumps by -1. Since $V-L$
varies by $+(D-2)$, then the genus $\tJ$ does not change. 

\item In the second subcase $i=n$ (or $i=m$): We have $D-2$ 
color pairs $kl \neq 0i$ which change by -1; $\widetilde F_{0im}$ changes by -1 if and only if $F_{0i}$ jumps by +1. This depends on the fact that 
the external lines of color $i$ of the dipole belong to a single
or to two external open $0i$ faces. In all cases, the genus 
does not change. 

\end{enumerate}

It remains to check the behavior of the boundary graph. 
 The modification here only involves the contraction of a regular melon 
of a $D$ dimensional color vacuum graph. This is a case completely
treated in the colored context (see \cite{GurRyan}) and it does not change the genus of any jacket. \qed

The following statement holds:
\begin{lemma}\label{lem3}
Cutting a white line in a rosette $\cG$,  the 
quantity $g_{\tJ} $ decreases, for any $\tJ$; the quantity $g_{\bJ}$
increases, for any $\bJ$, and at most by $D$. Hence, in particular, the quantity 
\bea
\sum_{J} g_{\tJ}  - k \sum_{\bJ} g_{\bJ}
\label{omed}
\eea
 decreases, for any $k\geq 0$. 
\end{lemma}
\noindent{\bf Proof.} Consider first a jacket $\tJ=(n0m\dots)$. 
Cutting a white line induces on the number vertices and lines 
\beq
V \to V \,, \qquad 
L \to L - 1\,, 
\eeq 
meanwhile for the faces of $\tJ$, $F_{kl}$
do not change (they do not pass by the 0-line cut);
there is 1 or 2 faces $F_{0n}$, $F_{0m}$ or
$\widetilde F_{nm0}$ which pass through the 0-line 
and they become, respectively, 2 faces or 1 face.
Hence,
\beq
F_{0n} + F_{0m}+\widetilde F_{nm0} \to 
F_{0n} + F_{0m}+\widetilde  F_{nm0} \pm 1\,.
\eeq
Thus,
\bea
g_{\tJ} \to g_{\tJ} \quad \quad \text{or}\quad \quad 
g_{\tJ} \to g_{\tJ} - 1 \,.
\eea
Let us now consider the boundary graph $\bG$, with a fixed boundary jacket $\bJ$.
Cutting the white line in $\cG$ changes the boundary $\bG$ into $\bG'$
(and $\bJ$ to $\bJ'$) with two new vertices $V_1$ and $V_2$
and $D$ new lines of color $1,\dots, D$ which can be partitioned
as a disjoint union $S\cup T$, where $S$ is the set of lines
joining the two new vertices.

Therefore we have $V_{\bG'} = V_{\bG} + 2 \, ,
\; L_{\bG'} = L_{\bG} + D$. If we define $\Delta c_\bG :=  c_{\bGa} - c_{\bG} $
(where $c_{\bG}$ denotes the number of connected components
of $\bG$)  
and $\Delta F_{\bJ} :=  F_{\bJ'} -F_{\bJ}$ then
\bea
g_{\bJ '}= g_{\bJ} + \Delta c_\bG  + \frac{D-2 - \Delta F_{\bJ}}{2}.
\label{ts}
\eea
In these notations the fact that $g_{\bJ}$ increases, and by at most $D$, is equivalent to 
\bea - D-2+ 2 \Delta  c_\bG  \le  \Delta F_{\bJ}  \le D-2 + 2 \Delta  c_\bG  . \label{boubou}
\eea

Given a pair of colors $ij$, we have
\bea
\begin{array}{c}
F_{ij} \to F_{ij} + 1 \qquad \text{if}\quad i,j \in S \,,\\
F_{ij} \to F_{ij} \qquad \text{if}\quad i \in S \quad j \in T 
\quad \text{(or vice-versa)} \,,\\
F_{ij} \to F_{ij} \pm  1 \qquad \text{if}\quad i,j \in T \,.
\end{array}
\label{fij}
\eea
The last case can be made more precise: we have $F_{ij} \to  F_{ij} -  1 $ exactly for those pairs $i$ $j$
for which the corresponding lines $i$ and $j$ in the boundary graph belonged to \emph{two distinct } $ij$ faces.

\begin{itemize}

\medskip\item 
If $T = \emptyset$ it is easy to check that $ \Delta c_\bG =1$ and $\Delta F_{\bJ} = D$, hence \eqref{boubou} is true.

\medskip\item 
If $S$ and $T$ are both non empty, in the new boundary graph the two new vertices $v_1$ and $v_2$ belong to a single
connected component (because they are necessarily joined by the lines of $S$). This component contains also all other end vertices of the $T$ chains. 
Observe that if these end vertices belonged to $p$ different connected components in the previous boundary graph,
so that $ \Delta c_\bG =1 - p $, there has to be also $p$ different components at least in the jacket cycle. Hence, at least $p-1$ pairs $ij$ in the considered jacket with $i \in T$, $j \in T$ which in the initial boundary graph belonged to \emph{two distinct } $ij$ faces. For them $F_{ij} \to  F_{ij} -  1 $. Furthermore, at least two
pairs of the jacket have $i \in S$, $j\in T$. Thus, carefully using \eqref{fij}
\bea
\Delta F_{\bJ} \le - (p-1) + [D- (2+ (p-1))]  = D- 2 p = D-2 + 2 \Delta c_\bG.
\eea 
Moreover again since at least two
pairs of the jacket have $i \in S$, $j\in T$, by \eqref{fij} we have $\Delta F_{\bJ} \ge - D +2$. Since $\Delta c_\bG -1 \le 0$
this more than implies \eqref{boubou}.

\begin{figure}
\centering
\begin{minipage}[t]{.7\textwidth}
\centering
\includegraphics[angle=0, width=7cm, height=2.75cm]{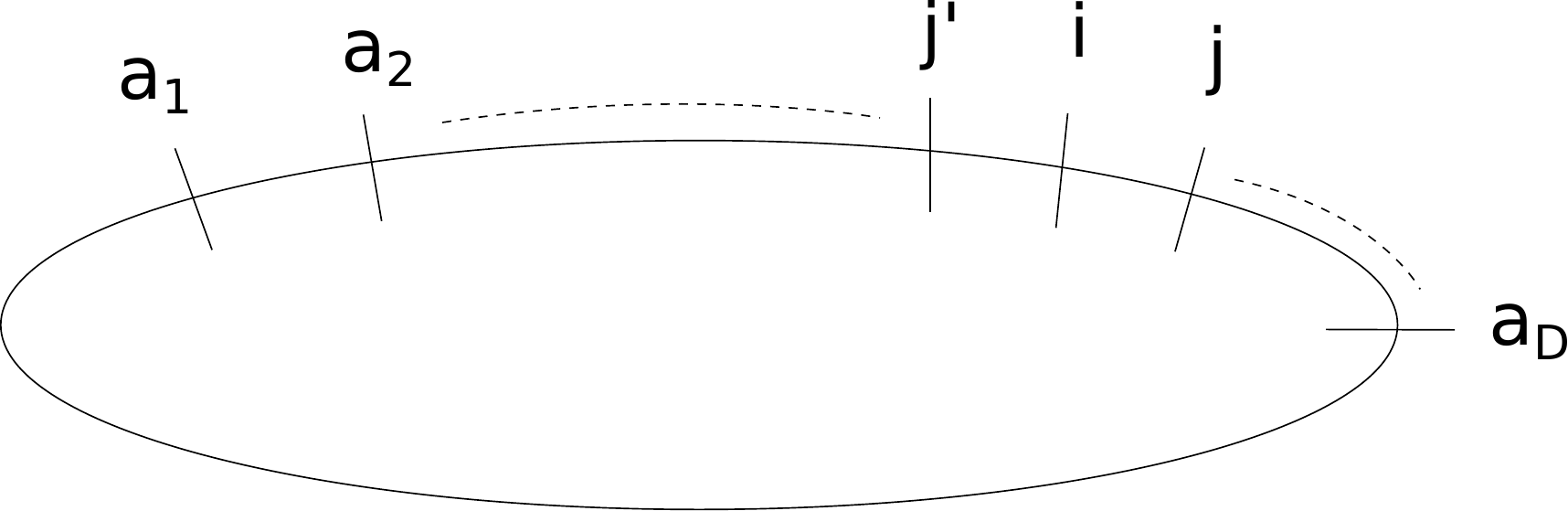}
\vspace{0.1cm}
\caption{ {\small Cycle of colors of the jacket. Each
color $i$ has two neighbors, $j$ and $j'$.}}
\label{fig:jc}
\end{minipage}
\end{figure}

\medskip\item 
If $S$ is empty, and the end vertices  of the $T$ chains belong
again to $p$ different components in $\bG$, they become either 1 or 2 components 
in $\bGa$ so that $1 - p \le \Delta c_\bG \le 2 -p$.
Since the $p$ different components break the \emph{cyclic} jacket ``circle" (see Figure \ref{fig:jc}) into at least $p$ different intervals, 
at least $p$ pairs in the  jacket cycle are associated with a change
of connected components. For any such pair we must have $F_{ij} \to  F_{ij} -  1 $. Therefore
\bea
\Delta F_{\bJ} \le - p + (D- p)  = D- 2 p \le D-2 + 2 \Delta c_\bG.
\eea 
Finally in this case $\Delta c_\bG \le 1$, and $\Delta F_{\bJ} \ge -D$, hence \eqref{boubou} holds again.

\end{itemize}
\qed

\section{Next-to-leading  amplitude analysis for a graph}

The analysis of the genera of the jackets in a graph with initial melonic vertices, such as those of \cite{bgriv},
can be first reduced, through a tree $T$ contraction followed by melonic contraction and using Lemmas \ref{lem1} and \ref{lem2} and Remark 2 above,
to the analysis of a vertex-melonic rosette without closed or open melopoles. If the initial graph was $D+1$ melonic, that rosette 
is empty. We carry out now this analysis in the opposite non-trivial case, for which we already remarked that the rosette
contains at least one non-melonic white line.

\medskip
\begin{figure}[h]
 \centering
     \begin{minipage}[t]{.7\textwidth}
      \centering
\includegraphics[angle=0, width=5.5cm, height=4cm]{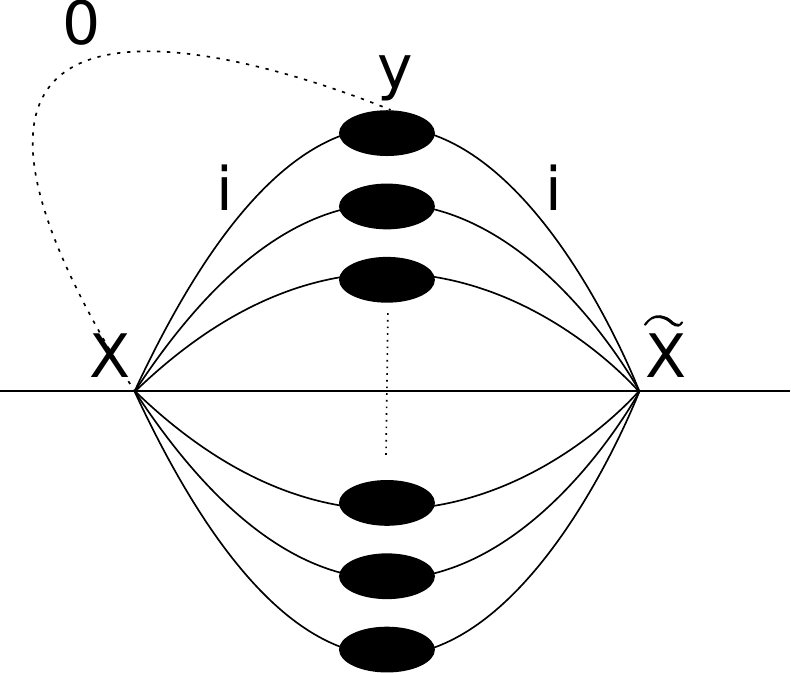}
\vspace{0.1cm}
\caption{ {\small A non-melonic white line (dotted 0-line) joining $x$ and $y\neq \tilde x$ yielding a subleading (non-melonic) graph.}}
\label{fig:xymelo}
\end{minipage}

\end{figure}

\begin{lemma}\label{lem4}
Let $\cG$ be a vertex-melonic rosette with a \emph{single non-melonic white line} $\ell$. We call $\cG'$ the graph in which we cut $\ell$. 
Then
\beq\label{k0deg}
\sum_{J} g_{\tJ} \geq \sum_{J} g_{\tJ'} + \frac{ D-2}{2}  (D-1) !
\eeq
and moreover
\beq\label{kdeg}
\sum_{J} g_{\tJ} - k \sum_{\bJ} g_{\bJ} \geq \sum_{J'} g_{\tJ'}- k \sum_{\bJ'} g_{\bJ'} + \frac{ D-2}{2}  (D-1) !
\eeq
\end{lemma}
\noindent{\bf Proof.} 
Since $\ell$ is non-melonic it joins two vertices $x$ and $y$ such that $y$ 
is not the mirror vertex  $\tilde x$ of $x$.  

It  trivially  follows from the Euler relation that, for ribbon graphs, cutting a single line decreases the genus by 0 or 1: 
\beq
V \to V \;, \qquad  L\to L-  1\,,  \qquad  F \to F\pm 1  \,.
\eeq
Now coming back to our present situation, 
 $y$ is on a single two-point function of color $i$ joining 
$x$ to $\tilde x$ (see Figure \ref{fig:xymelo}). Consider a pair $j,k$ of colors 
both distinct from $i$  and a jacket $\tJ = (j0k\dots)$. 
In this jacket, the graph $\cG'$ has two different faces of the 
$\widetilde F_{0jk}$-type, one touching $x$ and the other touching $y$. 
This is because any path from $x$ to $y$ not using $\ell$ must use color $i$, as seen in Figure \ref{fig:xymelo}. Indeed faces of the 
$\widetilde F_{0jk}$-type can use only lines of color j , k or white, and all white lines different from $\ell$
are melonic, hence all those touching the black melon containing $y$ must have their two ends in $y$.
Going from $\cG'$ to $\cG$ closes or merges these two 
faces in a single one. By the above remark on ribbon graphs, 
the genus of the jacket $\tJ$ must increase by 1.
Repeating the argument for all pairs $j,k$ distinct from 
$i$ and all jackets of the form $(j0k\dots)$ achieves the proof by a simple counting of all the jackets obtained in this way. 
Finally \eqref{kdeg} is a direct consequence of joining \eqref{k0deg}
and Lemma \ref{lem3}.

\qed

\section{Improved power counting for \cite{bgriv}}

Let us now specialize to $D=4$, and $k=4$ in Lemmas \ref{lem3}
and \ref{lem4}  and complete the proof of Proposition \ref{prop1}. 

Let $\cG$ be a graph of the theory considered in \cite{bgriv}  which satisfies the hypothesis
of Proposition 1, hence such that $\sum_{\bJ} g_{\bJ} = q\ge 1 $.
Its  successive reductions (contraction) by an arbitrary tree $T$ 
of white lines and then by melopoles is called $\cG_2 = \cG/(T\cup M)$. 
According to Lemmas \ref{lem1} and \ref{lem2} and since the initial vertices of $\cG$ were melons,
$\cG_2$ is a vertex-melonic rosette. It cannot be fully-melonic (as it would then have trivial melonic boundary) and, by Lemma \ref{lem3}, 
it has $r\ge q/12$ non-melonic lines, since 
cutting any non-melonic line moves each $ g_{\bJ}$ by at most  4 hence 
$\sum_{\bJ} g_{\bJ}$, at most, by 12.

Cutting $r-1$ non-melonic lines, one is led to a fully-melonic rosette $\cG_3$ with a single remaining non-melonic line. Applying Lemma \ref{lem4}, we obtain the following: 
\beq \label{fin1}
\sum_{J} g_{\tJ}- 4 \sum_{\bJ} g_{\bJ}  = \sum_{J_2} g_{\tJ_2} - 4 \sum_{{\bJ}_2} g_{{\bJ}_2}  \geq \sum_{J_3} g_{\tJ_3} - 4 \sum_{{\bJ}_3} g_{{\bJ}_3}   \ge  6\,.
\eeq
The last equality follows from the fact that $\cG_4 = \cG_3- \ell$ is fully melonic, 
hence  $\sum_{J} g_{\tJ_4} 
- 4 \sum_{\bJ} g_{{\bJ}_4} =0$.

This completes the proof of Proposition 1.
\qed
\medskip

\begin{figure}[h]
 \centering
     \begin{minipage}[t]{.7\textwidth}
      \centering
\includegraphics[angle=0, width=4cm, height=3cm]{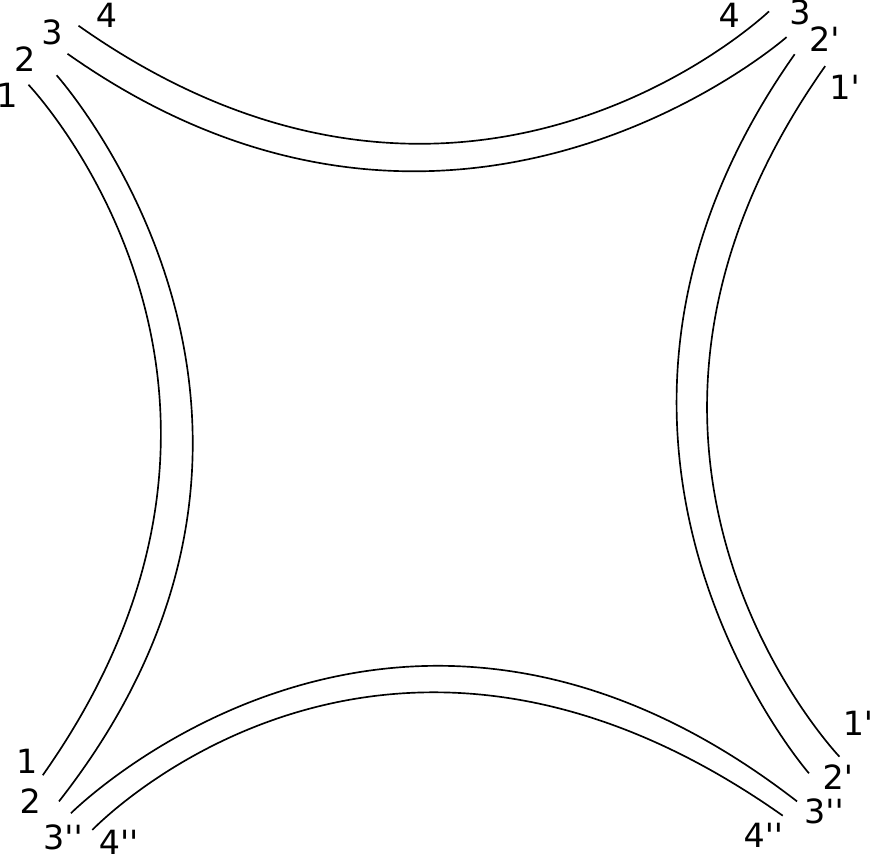}
\vspace{0.1cm}
\caption{ {\small $\phi^4$ matrix-like trace  invariant interaction.}}
\label{fig:mat4}
\end{minipage}

\end{figure}

Returning to the list in Section 5 of \cite{bgriv}, we remark that inequalities (82) and (85) were used to prove that graphs with $N_{\ext} =4$ and $\sum_{\bJ} g_{\bJ} = 1 $ were convergent. There is a single boundary graph for graphs of this type, namely the one
pictured in Fig. \ref{fig:mat4}.

Interestingly, this term could be interpreted as a $N^2$ by $N^2$ matrix-like 
invariant, generating a matrix-like sub-sector of the theory,
just like the anomaly term discussed in \cite{bgriv} generates a $N^4$ vector-like sector. It might be interesting in fact to 
construct models in which this subs-ector is divergent, as this may lead to richer models with vector, matrix and tensor aspects at once.

Nevertheless, this is not the case for the model treated in \cite{bgriv}. In that  model,
graphs with such an external structure converge and do not require any renormalization. Indeed, by Proposition 1 and Equality (42) of \cite{bgriv}\footnote{ We also use 
the present opportunity to correct a typo in the comment before Lemma 6,
in Equations (87) to (90)
and in the comment after (90) of \cite{bgriv}, $C_{\bJ}$ should be replaced instead by $C_{\bG}$.}
(putting $C_{\bG}=1, V_{4}=0=V_2= V_{2}''$), their divergence degree is at most $-2$. This concludes our analysis.

\section*{Acknowledgements}
The authors thank R. Gurau for useful discussions and thank F. Vignes-Tourneret for pointing out
a mistake in an earlier version of this paper.
Research at Perimeter Institute is supported by the Government of Canada through Industry
Canada and by the Province of Ontario through the Ministry of Research and Innovation.

\end{document}